\documentclass[11pt, a4paper]{article}
\pdfoutput=1
\usepackage{jcappub}

\usepackage{booktabs}
\usepackage{multirow}
\usepackage{amssymb}
\usepackage[export]{adjustbox}
\usepackage{floatrow}
\newfloatcommand{capbtabbox}{table}[][3.5in]
\newfloatcommand{capbfigbox}{figure}[][2.3in]
\usepackage{blindtext}
\usepackage{amsmath}
\usepackage{booktabs}
\usepackage{aas_macros}

\def\lsim{\mathrel{\raise.3ex\hbox{$<$\kern-.75em\lower1ex\hbox{$\sim$}}}}
\def\gsim{\mathrel{\raise.3ex\hbox{$>$\kern-.75em\lower1ex\hbox{$\sim$}}}}

\begin{document}

\hspace*{110mm}{\large \tt FERMILAB-PUB-18-528-A}

\vskip 0.2in

\title{Active Galactic Nuclei and the Origin of IceCube's Diffuse Neutrino Flux} 


\author{Dan Hooper$^{a,b,c}$,}\note{ORCID: http://orcid.org/0000-0001-8837-4127}
\emailAdd{dhooper@fnal.gov}
\author{Tim Linden$^d$}\note{ORCID: http://orcid.org/0000-0001-9888-0971}
\emailAdd{linden.70@osu.edu}
\author{and Abby Vieregg$^{c,e,f}$}\note{ORCID: http://orcid.org/0000-0002-4528-9886}
\emailAdd{avieregg@kicp.uchicago.edu}

\affiliation[a]{Fermi National Accelerator Laboratory, Theoretical Astrophysics, Batavia, IL 60510}
\affiliation[b]{University of Chicago, Department of Astronomy and Astrophysics, Chicago, IL 60637}
\affiliation[c]{University of Chicago, Kavli Institute for Cosmological Physics, Chicago, IL 60637}
\affiliation[d]{Ohio State University, Center for Cosmology and AstroParticle Physics (CCAPP), Columbus, OH  43210}
\affiliation[e]{University of Chicago, Department of Physics, Chicago, IL 60637}
\affiliation[f]{University of Chicago, Enrico Fermi Institute, Chicago, IL 60637}

\abstract{The excess of neutrino candidate events detected by IceCube from the direction of TXS 0506+056 has generated a great deal of interest in blazars as sources of high-energy neutrinos. In this study, we analyze the publicly available portion of the IceCube dataset, performing searches for neutrino point sources in spatial coincidence with the blazars and other active galactic nuclei contained in the Fermi 3LAC and the Roma BZCAT catalogs, as well as in spatial and temporal coincidence with flaring sources identified in the Fermi Collaboration's All-Sky Variability Analysis (FAVA). We find no evidence that blazars generate a significant flux of high-energy neutrinos, and conclude that no more than 5-15\% of the diffuse flux measured by IceCube can originate from this class of objects. While we cannot rule out the possibility that TXS 0506+056 has at times generated significant neutrino emission, we find that such behavior cannot be common among blazars, requiring TXS 0506+056 to be a rather extreme outlier and not representative of the overall blazar population. The bulk of the diffuse high-energy neutrino flux must instead be generated by a significantly larger population of less-luminous sources, such as non-blazar active galactic nuclei.}

\maketitle

\section{Introduction}

Beginning in 2013, the IceCube Collaboration has reported the observation of a diffuse flux of high-energy astrophysical neutrinos~\cite{Aartsen:2013bka,Aartsen:2015knd,Aartsen:2015rwa,Aartsen:2014gkd,Aartsen:2013jdh,Aartsen:2016xlq}. The spectrum of this emission is described by a power-law with an index of $\simeq$ 2.1-2.5 extending from tens of TeV to several PeV, and with measured flavor ratios that are consistent with those predicted from pion decay~\cite{Aartsen:2015ivb}. The angular distribution of this flux shows no significant departures from isotropy, and searches for point sources in the IceCube data have, until recently, produced only upper limits~\cite{Aartsen:2014cva,Aartsen:2015wto}, indicating that even the brightest sources of high-energy neutrinos contribute only a very small fraction of the total diffuse flux.

Since the discovery of this diffuse neutrino flux, significant effort has been directed toward identifying the sources of these very high-energy particles. Many possibilities have been proposed, including gamma-ray bursts~\cite{Waxman:1997ti,Piran:1999kx,Vietri:1998nm,Meszaros:2001ms,Guetta:2003wi}, star-forming galaxies~\cite{Loeb:2006tw}, and both blazar and non-blazar active galactic nuclei (AGN)~\cite{Stecker:1991vm,Halzen:1997hw,Atoyan:2001ey,Mannheim:1995mm,Muecke:2002bi} (for reviews, see Refs.~\cite{Becker:2007sv,Halzen:2002pg,Learned:2000sw,Halzen:1998mb}). Tidal disruption events~\cite{Wang:2015mmh,Senno:2016bso,Dai:2016gtz,Lunardini:2016xwi} and fast radio bursts~\cite{Aartsen:2017zvw,Fahey:2016czk} have also been discussed within this context recently. Most of these prospective source classes, however, have not held up to empirical scrutiny. The lack of neutrino candidate events observed in timing coincidence with gamma-ray bursts has been used to essentially rule out this class of objects as the primary source of IceCube's observed flux~\cite{Aartsen:2016qcr,Abbasi:2012zw,Aartsen:2018fpd} (low-luminosity gamma-ray bursts could potentially evade this constraint, however~\cite{Murase:2013ffa,Tamborra:2015qza,Senno:2015tsn}). On similar grounds, the lack of neutrino events spatially correlated with gamma-ray blazars appears to disfavor this source class as well~\cite{Aartsen:2016lir}. Furthermore, star-forming galaxies are also unable to generate this signal without exceeding the measured intensity of the isotropic gamma-ray background~\cite{Bechtol:2015uqb}.

The IceCube Collaboration recently announced the observation of an energetic muon track from the approximate direction of the blazar TXS 0506+056. This track was likely induced by a $\sim$300 TeV neutrino, and arrived during a period in which TXS 0506+056 was in a flaring state~\cite{IceCube:2018dnn}. The IceCube Collaboration estimates the statistical significance of this spatial and temporal coincidence to be approximately 3$\sigma$, after accounting for trials. Motivated by this potentially important event, the IceCube Collaboration subsequently performed a search of their full 9.5 year dataset in the direction of TXS 0506+056, identifying an additional excess of $13 \pm 5$ muon-neutrino events between September 2014 and March 2015, corresponding to a significance of 3.5$\sigma$~\cite{IceCube:2018cha}.

Taken at face value, this combination of results would appear to provide rather strong evidence in favor of blazars as a significant source of high-energy neutrinos (a na\"{i}ve combination of these two results yields a significance of 4.8$\sigma$). That being said, the multi-wavelength spectrum of TXS 0506+056, as measured by MAGIC, Fermi, Swift, and NuSTAR, does not reveal the presence of the cascade emission expected from sources capable of generating luminous neutrino flares. In particular, the flux that was observed from this object across the $\sim$$10^2-10^5$ eV range appears to rule out simple ({\it i.e.} single-zone) models in which the neutrinos are generated in the same region that is responsible for the observed electromagnetic emission~\cite{Keivani:2018rnh,Murase:2018iyl}. Although more complicated multi-zone models could potentially resolve this tension~\cite{Keivani:2018rnh,Murase:2018iyl,Liu:2018utd,He:2018snd}, the electromagnetic spectrum observed from TXS 0506+056 during its 2017 flaring state does not suggest that it is likely to be a significant source of high-energy neutrinos. Furthermore, TXS 0506+056 was not in an electromagnetic flaring state during the period of the observed neutrino excess in 2014-15~\cite{IceCube:2018cha}.

In this paper, we address the question of whether a significant fraction of the diffuse high-energy neutrino flux could originate from blazars, in a flaring state or otherwise. We utilize the one year of 86 string muon track data that has been made publicly available by the IceCube Collaboration (see \url{icecube.wisc.edu/science/data/PS-IC86-2011}). We perform searches for neutrino point sources in spatial coincidence with the blazars and other AGN contained in the Fermi 3LAC catalog~\cite{Ackermann:2015yfk} (building upon previous work by the IceCube Collaboration~\cite{Aartsen:2016lir}) and the Roma BZCAT blazar catalog~\cite{Massaro:2008ye}, as well as in spatial and temporal coincidence with the sources identified in the Fermi Collaboration's All-Sky Variability Analysis (FAVA)~\cite{2017ApJ...846...34A}. We find no evidence to support the conclusion that blazars generate high-energy neutrinos, and instead conclude that no more than 5-15\% of IceCube's diffuse flux can originate from this source class. While our analysis does not strictly rule out the possibility that the neutrino events associated with TXS 0506+056 do originate from this blazar, such a scenario would require this source to be a rather extreme outlier and not at all representative of members of known blazar populations. The bulk of IceCube's diffuse flux must instead be generated by a large population of much fainter sources, significantly more numerous than blazars. Within this context, we argue that non-blazar AGN are the most promising class of sources for the diffuse flux of high-energy neutrinos observed by IceCube.

\section{A Search For Neutrino Point Sources and Point Source Populations}
\label{method}

In this section, we describe our analysis of the muon track data released by the IceCube Collaboration for public use (\url{icecube.wisc.edu/science/data/PS-IC86-2011}). This dataset consists of muon tracks observed by IceCube between May of 2011 and May of 2012, which was the first full year that the detector was in its 86-string configuration. For each of the 138,322 neutrino candidate events included in this dataset, the direction and angular resolution is given, as well as a quantity called the ``energy proxy'', which roughly corresponds to the energy deposited in the detector. The dataset also includes the effective area of the detector as a function of declination and neutrino energy (averaged assuming an equal number of neutrinos and antineutrinos).

In our search for individual neutrino point sources, we follow an approach similar to that described in Ref.~\cite{Braun:2009wp} and implemented in Refs.~\cite{Aartsen:2014cva,Aartsen:2016oji}. For a dataset consisting of $N$ events, the probability density of event $i$ is given by:
\begin{equation}
\frac{n_s}{N} S_i + (1-\frac{n_s}{N}) B_i,
\end{equation}
where $n_s$ is the number of events originating from the point source and $S_i$ and $B_i$ are the signal and background probability distribution functions (PDFs), respectively. The likelihood of the data given $n_s$ is the product of these probability densities:
\begin{equation}
\mathcal{L}(n_s) = \prod^N_{i=1} \bigg[\frac{n_s}{N} S_i + (1-\frac{n_s}{N}) B_i\bigg].
\label{like}
\end{equation}
To determine the statistical significance in favor of a neutrino point source in a given detection of the sky, we calculate the change in the log-likelihood, as evaluated for the value of $n_s$ that provides the best fit to the data, $\Delta \ln \mathcal{L} = \ln \mathcal{L}(\hat{n}_s)-\ln \mathcal{L}(0)$.

%

The signal PDF for event $i$ is given as follows:
\begin{equation}
S_i = \frac{1}{2 \pi \sigma^2_i} \, e^{-\frac{|\vec{x}_i - \vec{x}_s|^2}{2\sigma^2_i}}, 
\end{equation}
where $\vec{x}_s$ is the direction of the hypothesized source, $\vec{x}_i$ is the reported direction of the event, and $\sigma_i$ is the uncertainty associated with the arrival direction of the event. The background PDF, $B_i$, is determined empirically, and is set to the number of events in the dataset per solid angle averaged across a band of $\pm 6^{\circ}$ in declination around the source in question (following Refs.~\cite{Aartsen:2014cva,Aartsen:2016oji}). Due to the insufficient solid angle available to characterize the background near the poles, we only consider sources with declinations between $+87^{\circ}$ and $-87^{\circ}$.

Up to this point in our analysis, there are two ways in which our point-source search differs from those presented in Refs.~\cite{Aartsen:2014cva,Aartsen:2016oji}. First, we have access to only one year of 86-string data, whereas Ref.~\cite{Aartsen:2014cva} makes use of a larger dataset collected over a longer period of time. Second, we do not directly utilize any information associated with the energy of each event. Although the IceCube public data release does include a quantity called the ``energy proxy'' for each event, it is not clear how this is precisely defined or how to quantitatively incorporate it into our analysis without having access to simulations of IceCube's instrumental response. The spectral shape of the neutrino flux impacts our analysis only through the energy dependence of IceCube's effective area.

\begin{figure}
\includegraphics[width=3.0in,angle=0]{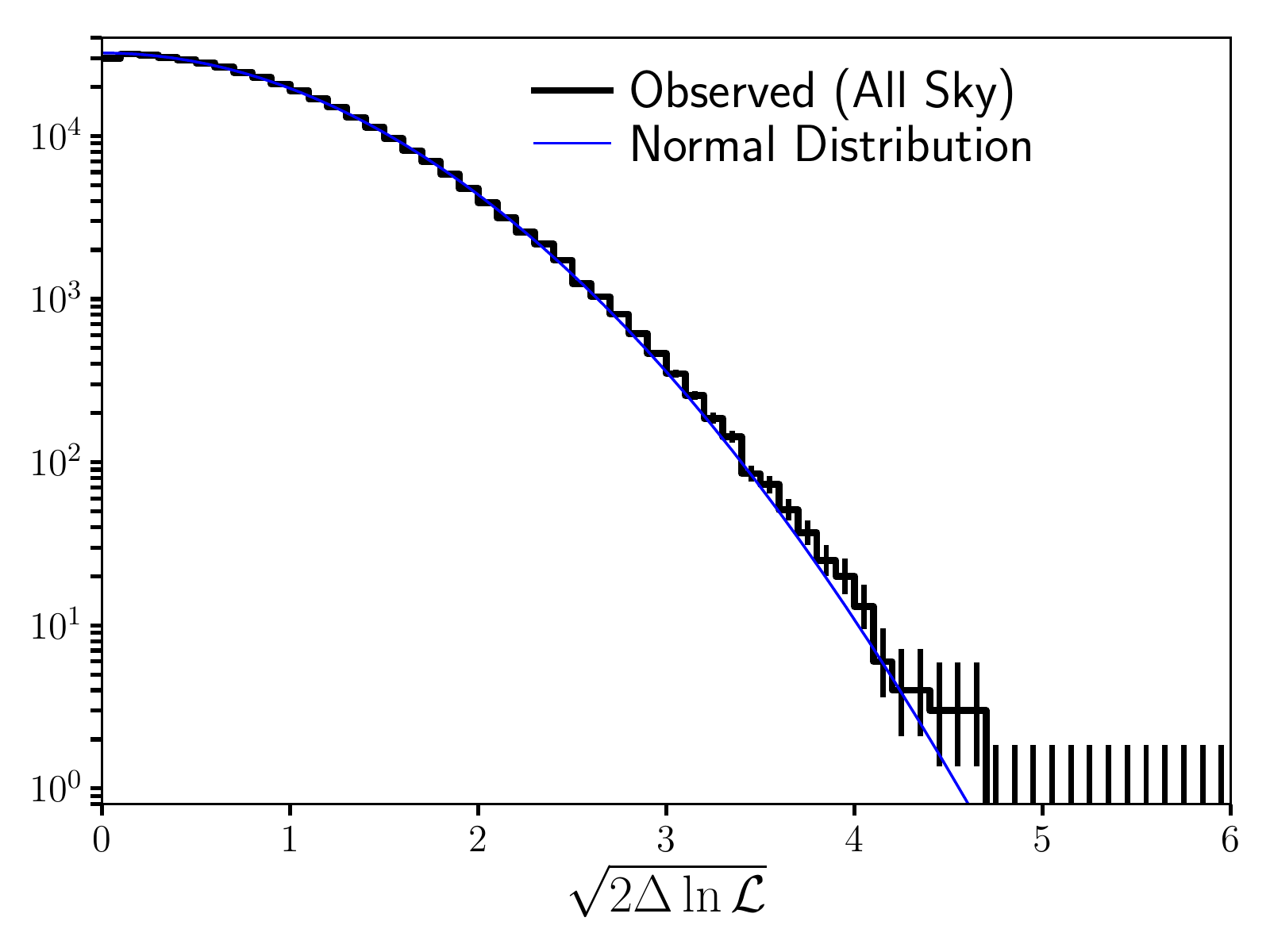}
\caption{The likelihood distribution in favor of a neutrino point source at locations across the sky in a grid of $0.2^{\circ}$ steps. The observed distribution shows no compelling evidence of any neutrino point sources. Sky locations with $\Delta \ln \mathcal{L} < 0$ (corresponding to a best fit with a negative point source flux) are not shown. We also include error bars conveying the 68\% Poissonian confidence interval on each bin.}
\label{histscan}
\end{figure}

\begin{figure}
\includegraphics[width=4.0in,angle=0]{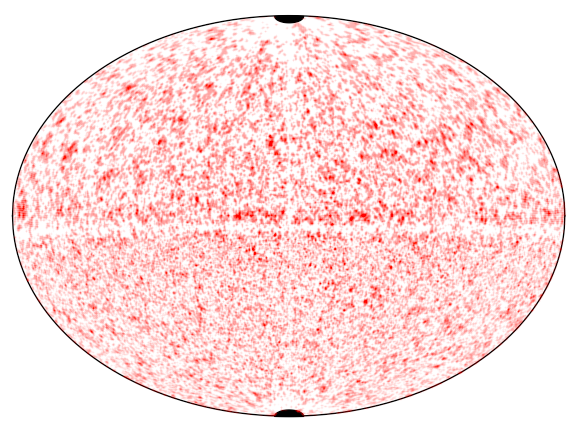}
\caption{A map of the evidence in favor of a neutrino point source, $2\Delta \ln \mathcal{L}$, in RA and Dec (in Aitoff projection). The darkness of the points scale linearly with $2\Delta \ln \mathcal{L}$, from 0 (white) to 22 (darkest red).}
\label{map}
\end{figure}

As a first test for the presence of neutrino point sources in the IceCube dataset, we scanned over the entire sky in a grid, taking $0.2^{\circ}$ steps, and calculating the $\Delta \ln \mathcal{L}$ in favor of a neutrino point source at each location. This test identified no compelling evidence of any point source population, and we show in Fig.~\ref{histscan} that the observed likelihood distribution is in reasonable agreement with that predicted from Gaussian fluctuations of the background. Although there appears to be a slight excess of sky locations with $\sqrt{2\Delta \ln \mathcal{L}} \sim 3-5$, we consider it entirely plausible that this arises from small inaccuracies in our PDFs. The most statistically significant sky locations each yield $\sqrt{2\Delta \ln \mathcal{L}} \approx 4.7$, consistent with background expectations in light of the large trials factor associated with the all-sky scan (and consistent with previous studies that find no evidence for point sources in this dataset~\cite{Aartsen:2014cva,Aartsen:2016oji}). In Fig.~\ref{map} we present a map of the values of $2\Delta \ln \mathcal{L}$ found in our analysis.

Next, we used the IceCube dataset to search for evidence of a population of neutrino point sources associated with known classes of astrophysical objects, such as blazars and other AGN. In doing so, we considered three different hypotheses for the expected neutrino emission from the members of a given source class:
\begin{enumerate}
\item{Gamma-Ray Scaling: The neutrino flux from a given source is proportional to the gamma-ray flux observed from that source, $n_s \propto F_{\gamma}$ (in units of photons between 1-100 GeV per area, per time as reported in the 3FGL catalog~\cite{Acero:2015hja}, unless stated otherwise). This hypothesis would be valid in cases in which the observed gamma-ray emission is produced mostly through hadronic interactions, yielding both charged and neutral pions and thus a fixed ratio of neutrinos and photons.}
\item{Flat Scaling: The neutrino flux of each source is uncorrelated to any other information (other than membership in the catalog under consideration), corresponding to equal values of $n_s$ for all sources in the catalog.}
\item{Geometrical Scaling: The neutrino flux of a given source is proportional to $1/D_L^2$, where $D_L$ is the luminosity distance of the source, $n_s \propto D_L^{-2}$.\footnote{We relate the luminosity distance of a source to its redshift through the following: $D_L = \frac{c(1+z)}{H_0} \int_0^z  dz' [\Omega_M(1+z')^3+\Omega_{\Lambda}]^{-0.5}$, adopting the best-fit values of $H_0$, $\Omega_M$ and $\Omega_{\Lambda}$ as reported by the Planck Collaboration~\cite{Aghanim:2018eyx}.} This hypothesis treats the neutrino luminosity of a given source as uncorrelated to other information, taking only into account the distance between the source and observer. Note that this approach can be applied only to those sources with measured redshifts, and thus in some cases require us to utilize smaller source catalogs.}
\end{enumerate}

The first two of these hypotheses were previously applied in Ref.~\cite{Aartsen:2016lir}, in which the IceCube Collaboration performed a search for neutrino point sources among the blazars in Fermi's 2LAC catalog. We build upon this work by also considering the geometrical scaling hypothesis, and by making use of the significantly larger 3LAC~\cite{Ackermann:2015yfk} and Roma-BZCAT~\cite{2015Ap&SS.357...75M} catalogs. We also consider blazar flares (as identified in the FAVA catalog~\cite{2017ApJ...846...34A}) that took place during the time period of IceCube's dataset, as well as non-blazar source classes.

The joint likelihood in each case is calculated as the product of the likelihoods for each source, as described in Eq.~\ref{like}, as a function of the total neutrino flux from the entire source population. From this exercise, it is possible to either identify evidence of neutrino emission from a given collection of sources, or to set limits on the total neutrino flux from the source catalog under consideration.

\begin{figure}
\includegraphics[width=3.0in,angle=0]{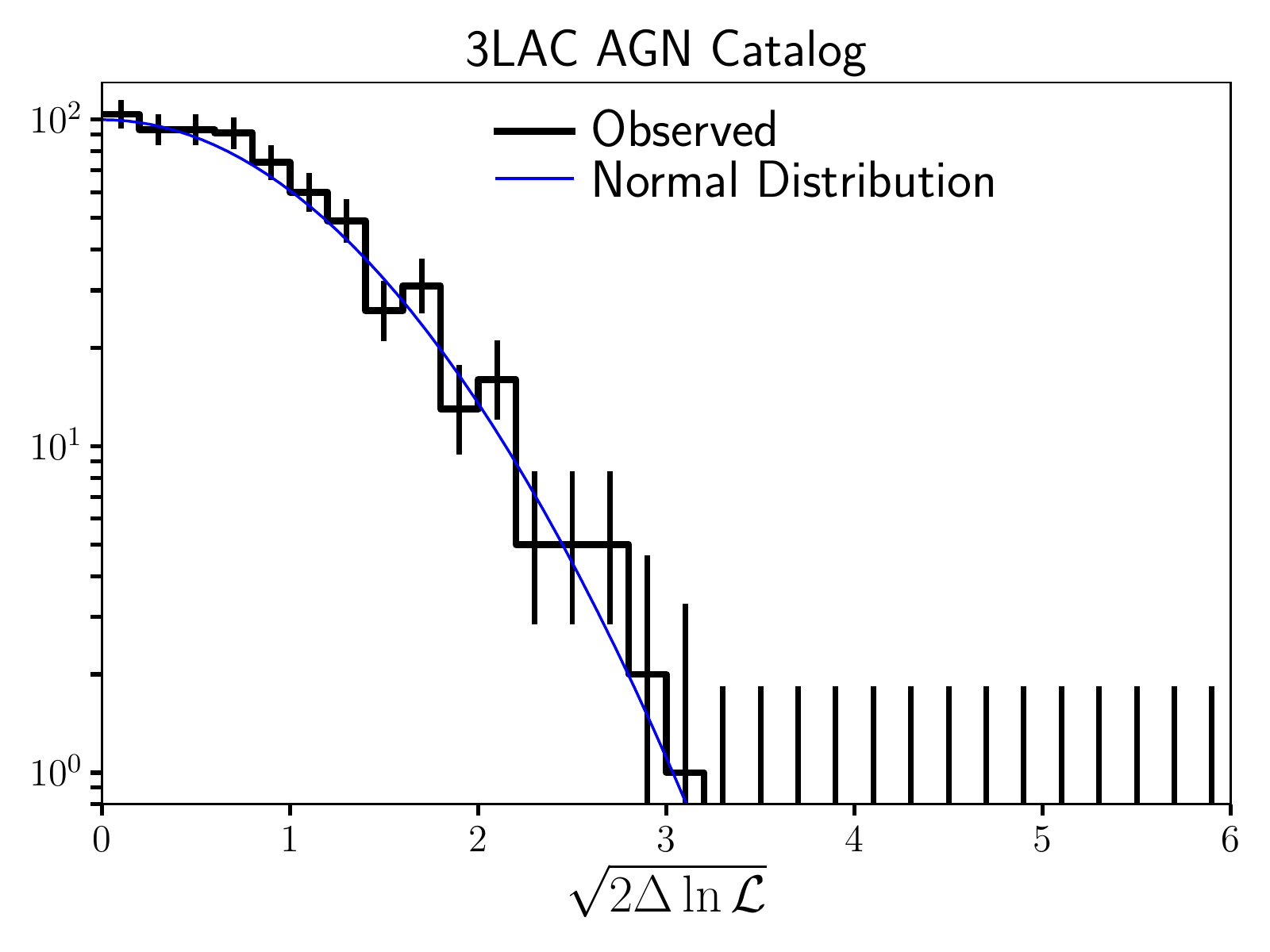} \caption{The likelihood distribution in favor of a neutrino point source in the directions of 1769 AGN contained in Fermi's 3LAC catalog~\cite{Ackermann:2015yfk}. The measured distribution shows no evidence of neutrino emission from this collection of sources.}
\label{hist3LAC}
\end{figure}

\section{Neutrinos from 3LAC Blazars}

We begin by considering 1769 of the 1773 AGN contained in the Fermi Collaboration's 3LAC catalog, of which 896 have reported redshifts~\cite{Ackermann:2015yfk} (four of the 3LAC sources were not used in our analysis due to their locations near the poles). Although this catalog contains other types of AGN, it is dominated by blazars and we use it to test whether the diffuse neutrino flux observed by IceCube might originate entirely, or in part, from this class of sources. In Fig.~\ref{hist3LAC}, we plot the likelihood distribution in favor of neutrino point sources at the locations of these 1769 sources. This is well described by a normal distribution, showing no indication of neutrino emission from this class of objects. In Table~\ref{table1}, we list each of the 11 AGN in the 3LAC catalog that yielded $2\Delta\ln \mathcal{L}>6$, along with their redshift (if known) and gamma-ray flux as measured by Fermi~\cite{Acero:2015hja}. We are unable to identify anything about this collection of sources that sets them apart from other representative samples of sources in the 3LAC catalog.

\begin{table*}
\label{Table1}
\begin{tabular}{|c|c|c|c|c|c|c|}
\hline 
3FGL Name  & Other Name(s)   & Type  & $z$ & $\Phi_{1-100 \,{\rm GeV}}$ & $2\Delta \ln \mathcal{L}$ \tabularnewline
  &   &   & &  (cm$^{-2}$ s$^{-1}$) &  \tabularnewline
\hline 
\hline 
J2235.3-4835 & PKS 2232-488     & FSRQ   & 0.51 &  $7.45 \times 10^{-10}$   & 10.2 \tabularnewline
                                 & 5BZQJ2235-4835 &  & & &  \tabularnewline
\hline
J2152.9-0045 & RBS 1792 & BL Lac  & 0.341 & $3.38 \times 10^{-10}$  & 8.60 \tabularnewline
&5BZBJ2153-0042 &&&&   \tabularnewline
\hline
J0358.7+0633 & PMN J0358+0629  & Unknown  & -- & $6.02 \times 10^{-10}$  & 8.32 \tabularnewline
\hline
J1016.0+0513 & TXS 1013+054 & FSRQ & 1.714  &  $2.52 \times 10^{-9}$  & 7.78 \tabularnewline
& 5BZQJ1016+0513 & &&& \tabularnewline
\hline
J0658.6+0636 & NVSS J065844+063711 & Unknown & -- & $5.07 \times 10^{-10}$  & 7.24 \tabularnewline
\hline
J2039.0-1047 & TXS 2036-109  & BL Lac& -- & $2.39 \times 10^{-9}$ & 7.21\tabularnewline 
& 5BZBJ2039-1046 & &&& \tabularnewline
\hline
J0353.0-3622 & XRS J0353-3623 &BL Lac & -- & $1.99 \times 10^{-10}$& 6.97 \tabularnewline
& 5BZBJ0353-3623 & &&& \tabularnewline
\hline
J1018.5+0530 & TXS 1015+057 & FSRQ & 1.944 & $7.84 \times 10^{-10}$ & 6.77 \tabularnewline
&  5BZQJ1018+0530 & &&& \tabularnewline
\hline
J1251.3+1041 & 1RXS J125117.4+103914 &  BL Lac & 0.2454 & $ 3.69 \times 10^{-10}$ & 6.73 \tabularnewline
&  5BZBJ1251+1039   & &&& \tabularnewline
\hline
J1146.8+3958 & NVSS J114653+395751 & Unknown  & --  & $ 3.34\times 10^{-9}$ & 6.23 \tabularnewline
\hline
J1516.9+1926 & PKS 1514+197   & BL Lac & -- & $ 2.80 \times 10^{-10}$ & 6.12 \tabularnewline
&  5BZBJ1516+1932 & &&& \tabularnewline
\hline
\hline
\end{tabular}
\caption{Each of the 11 AGN in the 3LAC catalog~\cite{Ackermann:2015yfk} that yielded evidence for neutrino emission at a level of $2\Delta\ln \mathcal{L}>6$, along with their redshift (if known) and gamma-ray flux as measured by Fermi~\cite{Acero:2015hja}.}
\label{table1}
\end{table*}

\begin{figure}
\includegraphics[width=3.0in,angle=0]{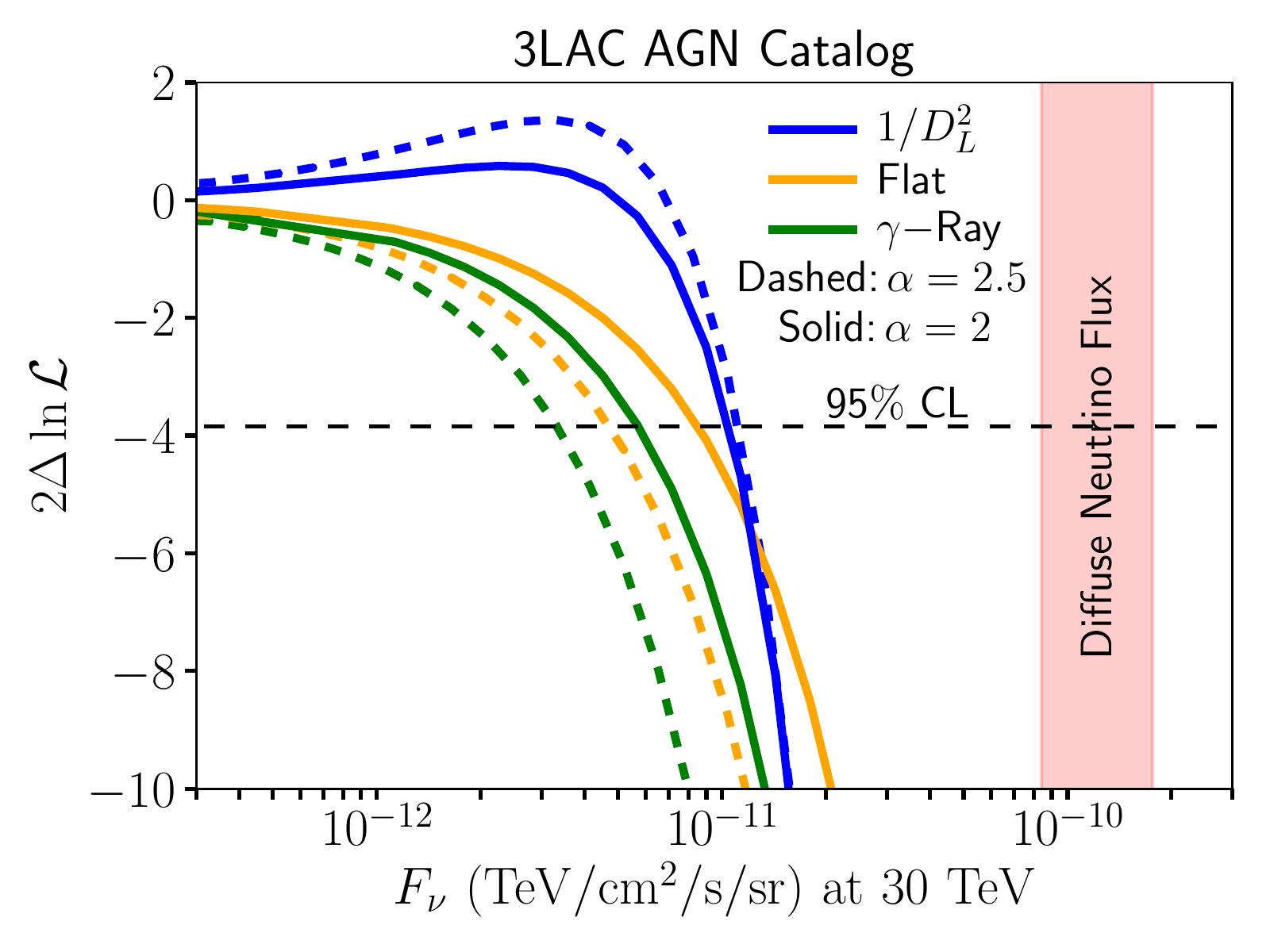}
\includegraphics[width=3.0in,angle=0]{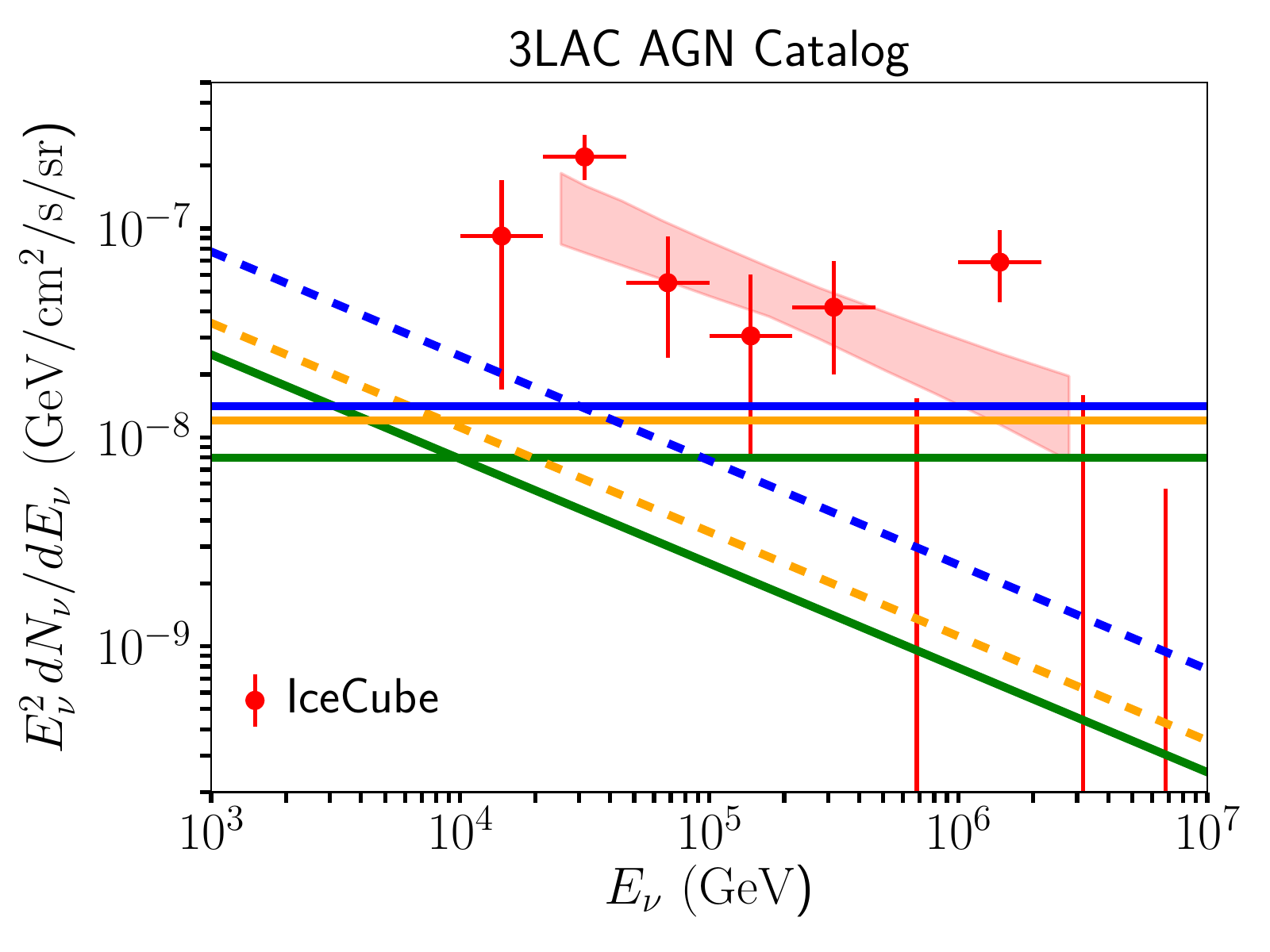}
\caption{Left frame: The change in the log-likelihood as a function of the total, all-flavor, neutrino emission from the AGN contained in the 3LAC catalog, for two choices of the neutrino spectral index, $\alpha$, and for the three scaling hypotheses described in Sec.~\ref{method} ($1/D_L^2$, Flat, $\gamma$-Ray). We do not obtain any statistically significant evidence for neutrino emission from this source class, and in each case we place a 95\% confidence level limit on the total neutrino emission that is approximately an order of magnitude below the diffuse neutrino flux reported by the IceCube Collaboration~\cite{Aartsen:2015knd} (see also Refs.~\cite{Aartsen:2016xlq,Aartsen:2015rwa,Aartsen:2015knd}). Right frame: Upper limits on the total flux of neutrinos from the sources in the 3LAC catalog, multiplied by 1.4 to account for the emission from blazars that are too gamma-ray faint to be included in this catalog, and using the same line and color legend as in the left frame. No more than $\simeq$\,5-15\% of the diffuse neutrino flux measured by IceCube can be generated by blazars.}
\label{3LAC}
\end{figure}

In Fig.~\ref{3LAC}, we present the results of our population analysis as applied to the members of the 3LAC catalog. In the left frame of this figure, we plot the value of $2\Delta\ln \mathcal{L}$ obtained in our analysis as a function of the total neutrino emission from this collection of sources (evaluated at an energy of 30 TeV), for each the three scaling hypotheses described in the previous section of this paper. We also adopt a power-law spectrum ($dN_{\nu}dE_{\nu} \propto E_{\nu}^{-\alpha}$) with an index of either $\alpha=2.0$ or 2.5. In none of these cases do we obtain any statistically significant evidence for neutrino emission and we proceed to place a 95\% confidence level upper limit (corresponding to $2\Delta \ln \mathcal{L} = -3.84$) on the total neutrino emission from this collection of sources. The limits we obtain are approximately an order of magnitude below the diffuse flux reported by the IceCube Collaboration~\cite{Aartsen:2015knd} (see also Refs.~\cite{Aartsen:2016xlq,Aartsen:2015rwa,Aartsen:2015knd}). 

In the right frame of this figure, we plot these limits compared directly to the diffuse neutrino flux reported by the IceCube Collaboration~\cite{Aartsen:2015knd}, using the same line and color legend as in the left frame. In this frame, however, we have multiplied the limit by a factor of 1.4 to account for the emission from blazars that are too distant and/or too low in gamma-ray luminosity to be included in the 3LAC catalog (the same factor was included in the analysis of Ref~\cite{Aartsen:2016lir}, and we are being slightly conservative in applying it to the case of the larger 3LAC catalog). From this comparison, we conclude that no more than $\simeq$\,5-15\% of IceCube's neutrino flux at 30 TeV can be generated by blazars, although we cannot rule out the possibility that a larger fraction of the flux at higher energies originates from this class of sources. While these results are consistent with those presented by the IceCube Collaboration~\cite{Aartsen:2016lir}, we note that our analysis benefits from the use of the 3LAC catalog, which contains approximately 70\% more AGN than were utilized in the analysis by the IceCube Collaboration.

\begin{figure}
\includegraphics[width=3.0in,angle=0]{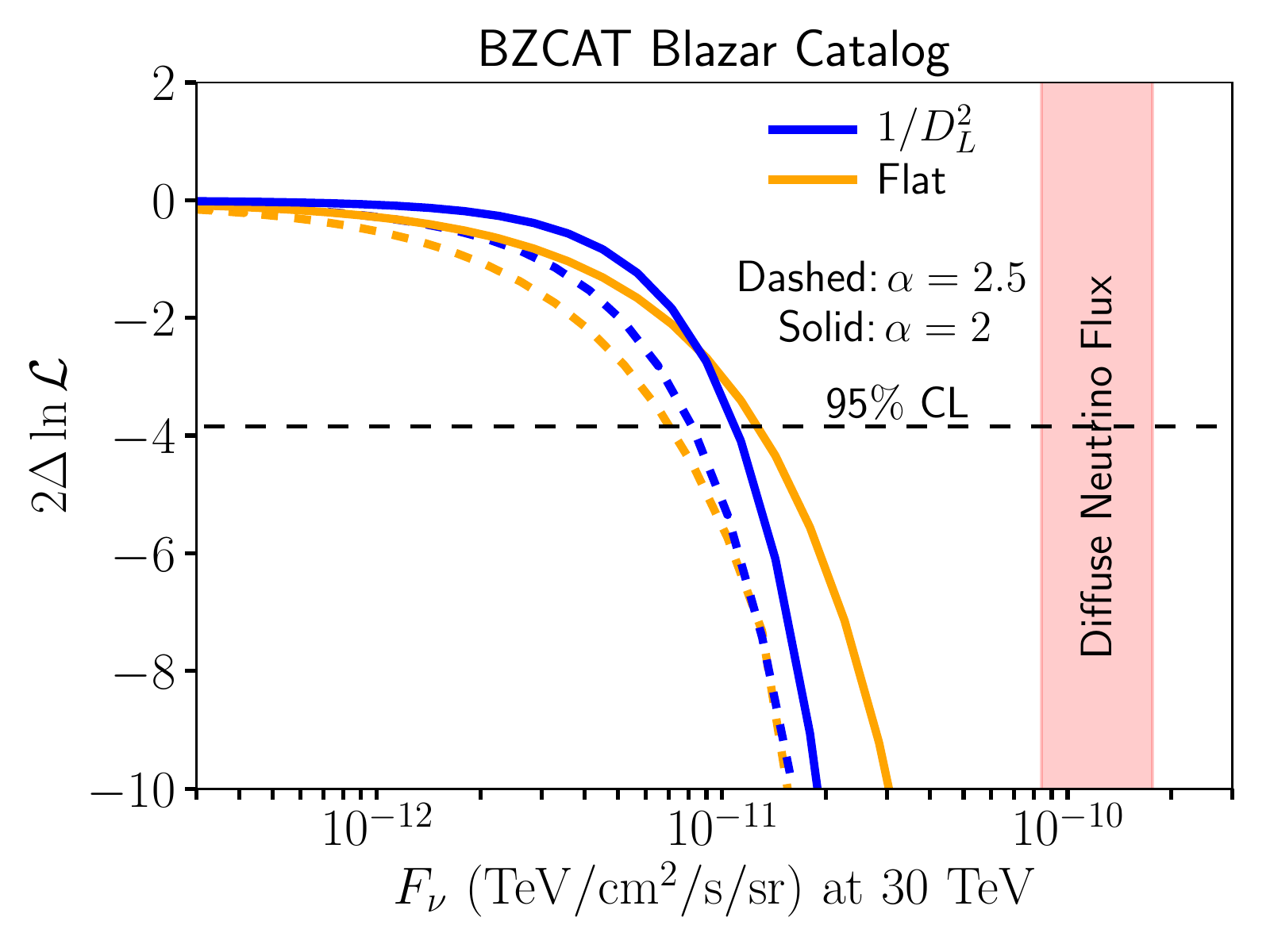} 
\caption{As in Fig.~\ref{3LAC}, but for the case of blazars contained in the Roma-BZCAT catalog.}
\label{bzcat}
\end{figure}

In each case shown in the right frame of Fig.~\ref{3LAC}, we applied the same factor of 1.4 to account for the blazars that are not members of the 3LAC catalog. This value is based on the total gamma-ray intensity from unresolved blazars, and it is not clear that this is the correct quantity to apply in the cases of the flat scaling or geometric scaling hypotheses. As an extreme illustration of this point, one could imagine a scenario in which the neutrino emission from a given blazar is anti-correlated to its gamma-ray emission. In this case, a large fraction of the total neutrino emission from this class of sources could originate from blazars that are not contained in the 3LAC catalog~\cite{Murase:2015xka}. With this in mind, we have repeated this exercise using the significantly larger Roma-BZCAT Multifrequency Catalog of Blazars~\cite{2015Ap&SS.357...75M}, which contains 3561 sources (including 2842 with reported redshifts) that are either confirmed blazars or exhibit blazar-like characteristics.  In Fig.~\ref{bzcat} we plot the results of this analysis, finding again no evidence of neutrino emission and allowing us to conclude that gamma-ray faint blazars do not substantially contribute to IceCube's diffuse neutrino flux.

Before moving on, we note that the blazar TXS 0506+056 shows no indication of any neutrino emission in this dataset ($2\Delta \ln \mathcal{L}<0$ for this source direction). In particular, this dataset includes 12 (47) events with best-fit arrival directions that are within $1^{\circ}$ ($2^{\circ}$) of the location of TXS 0506+056, entirely consistent with the expected background of 13.0 (52.1) events. However, it should be kept in mind that the detections of this source reported by the IceCube Collaboration are based on observations taken over different periods of time and thus are not incompatible with this null result.

The analysis procedure described above is designed to produce reliable constraints on the neutrino flux from a given source catalog. In particular, the flat-scaling case is predicted to yield a constraint that is valid for {\it any} distribution of neutrino fluxes from the collection of sources under consideration (for discussion, see Ref.~\cite{Aartsen:2016lir}). That being said, it is possible that the statistical significance in favor of a neutrino flux from a collection of sources might be reduced if we are not testing a scaling hypothesis that is sufficiently similar to that actually realized in nature. For example, one could imagine a scenario in which there is, on average, a proportionality between the neutrino and gamma-ray emission from a given AGN, but with a large degree of scatter from source-to-source. With this in mind, we introduce the following likelihood function:\footnote{For those readers familiar with gamma-ray searches for dark matter annihilation products in dwarf galaxies, the quantities $F_j$ and $\delta_j$ in the likelihood function are analogous to the $J$-factors and their uncertainties~\cite{Ackermann:2013yva}.}
\begin{equation}
\tilde{\mathcal{L}} = \prod_j \mathcal{L}_j(n_s) \times \frac{1}{\ln(10) F_j \sqrt{2\pi} \delta} \, e^{-(\log_{10} F_j-\overline{\log_{10} F_j})^2/2\delta^2},
\label{delta}
\end{equation}
where the sum is over the list of sources, $F_j$ is the neutrino flux from source $j$, $\overline{\log_{10} F_j}$ is the logarithm of the neutrino flux predicted from the proportionality to the observed gamma-ray flux, and $\delta$ is the variance in the neutrino-to-gamma ray flux ratio, in decades. We treat $F_j$ as nuisance parameters, integrating over them to yield a likelihood that is a function of only the mean neutrino-to-gamma ray ratio.

\begin{figure}
\includegraphics[width=3.0in,angle=0]{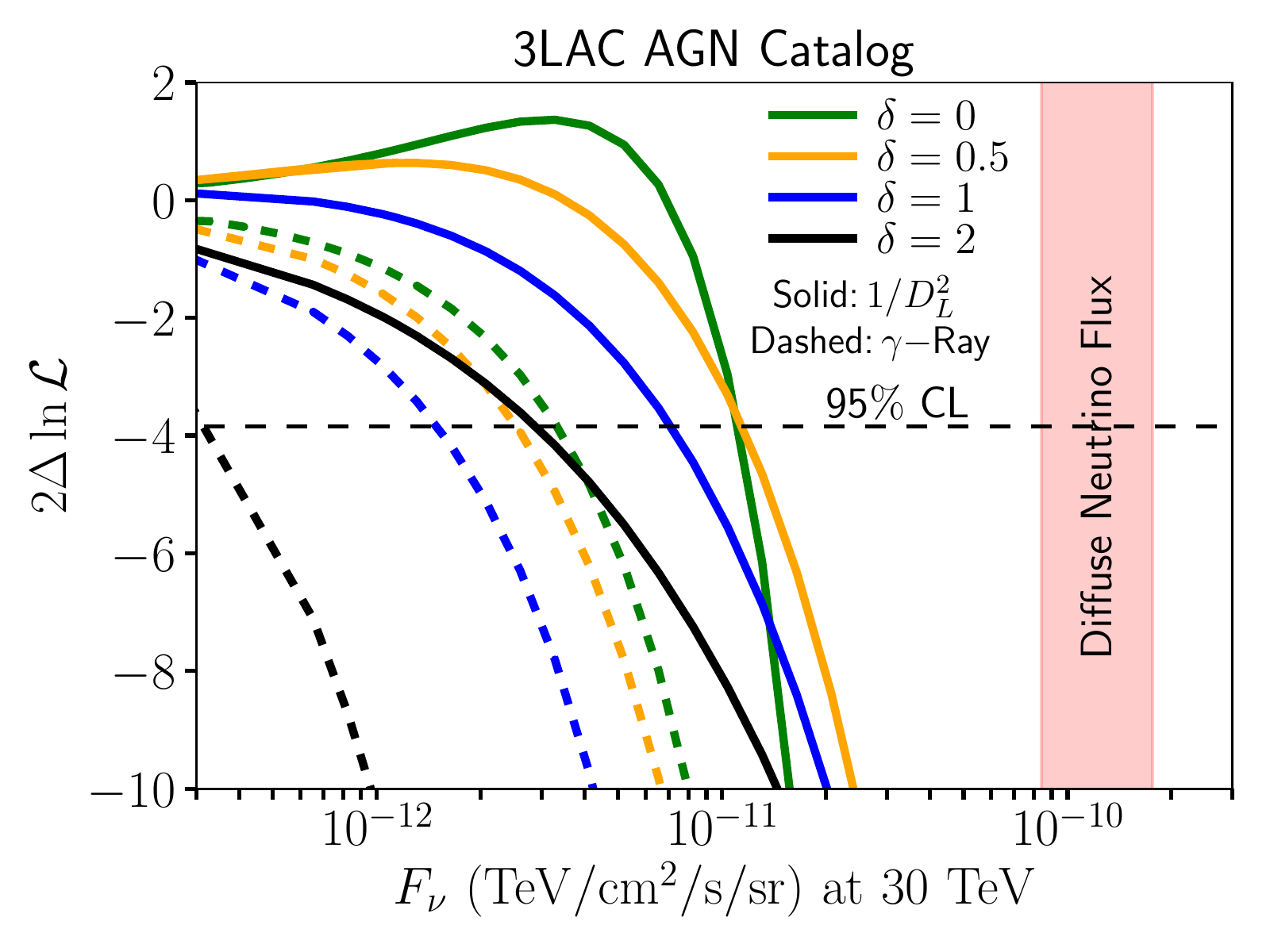} 
\caption{As in Fig.~\ref{3LAC}, but for a spectral index of $\alpha=2.5$ and for four values of $\delta$ (see Eq.~\ref{delta}). Considering source-to-source variations in the neutrino emission from blazars only strengthens the limits that can be placed on the total neutrino flux from this source population.}\label{TSsigma}
\end{figure}

\begin{figure}
\includegraphics[width=3.0in,angle=0]{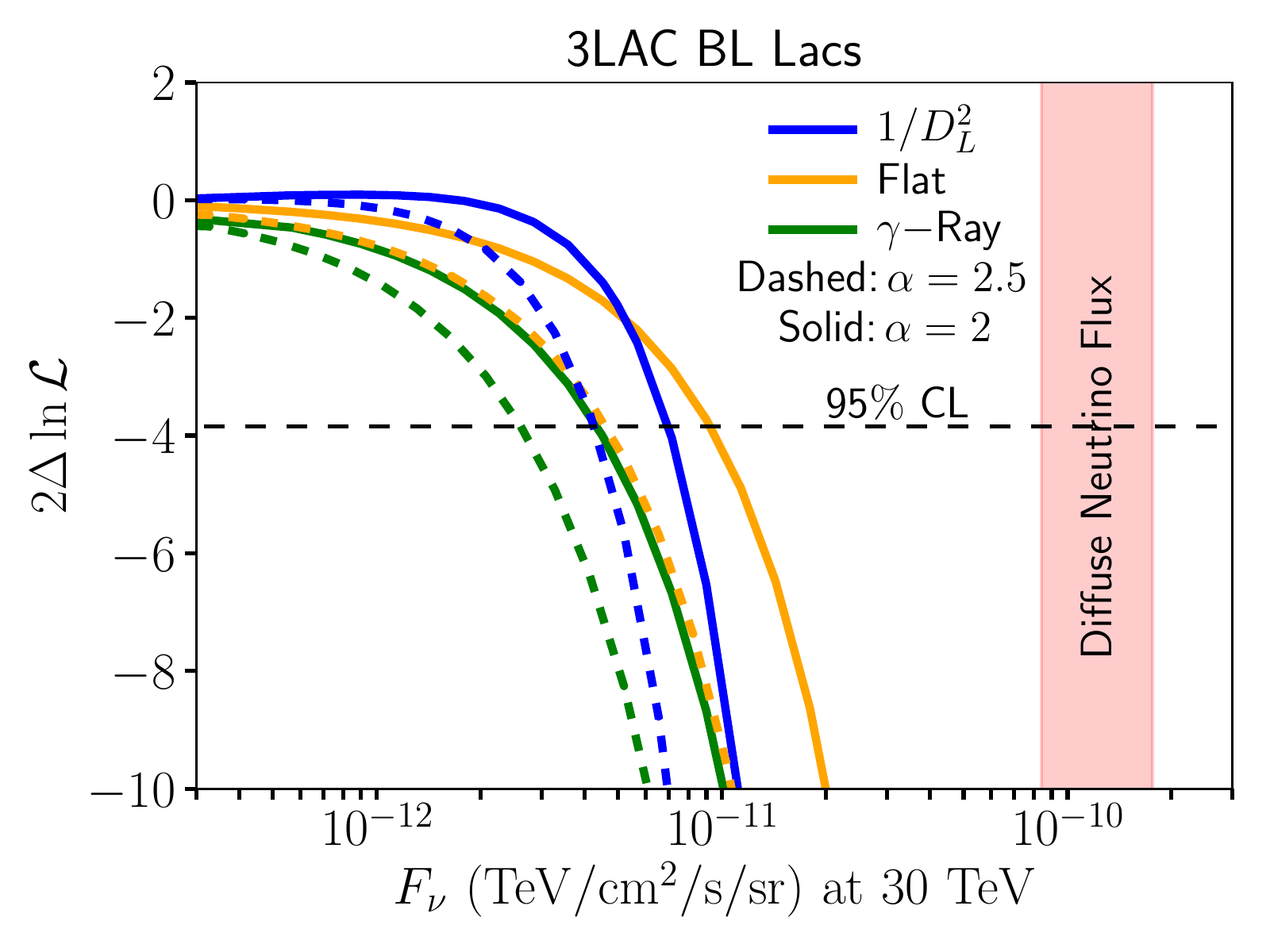}
\includegraphics[width=3.0in,angle=0]{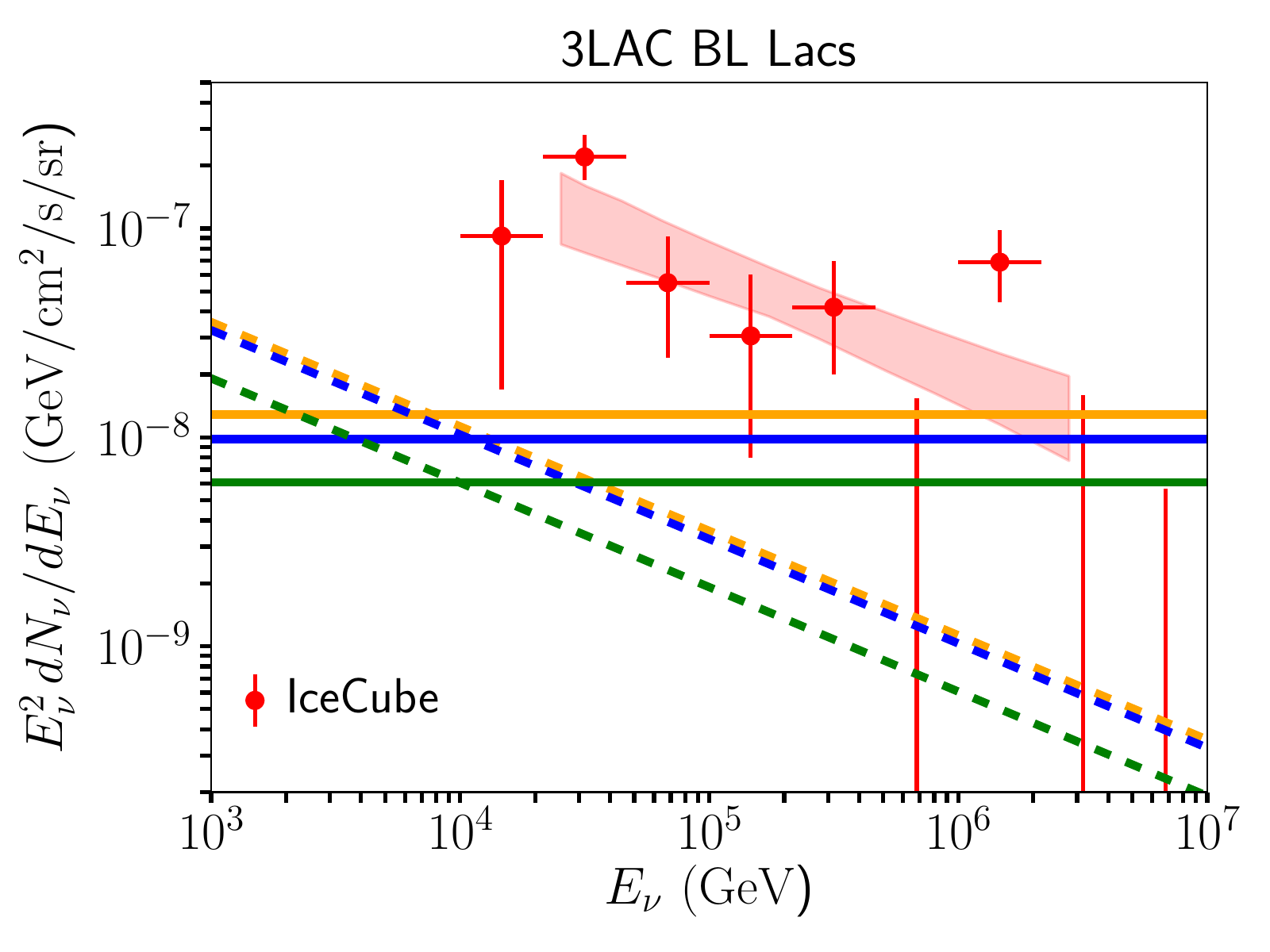} \\
\includegraphics[width=3.0in,angle=0]{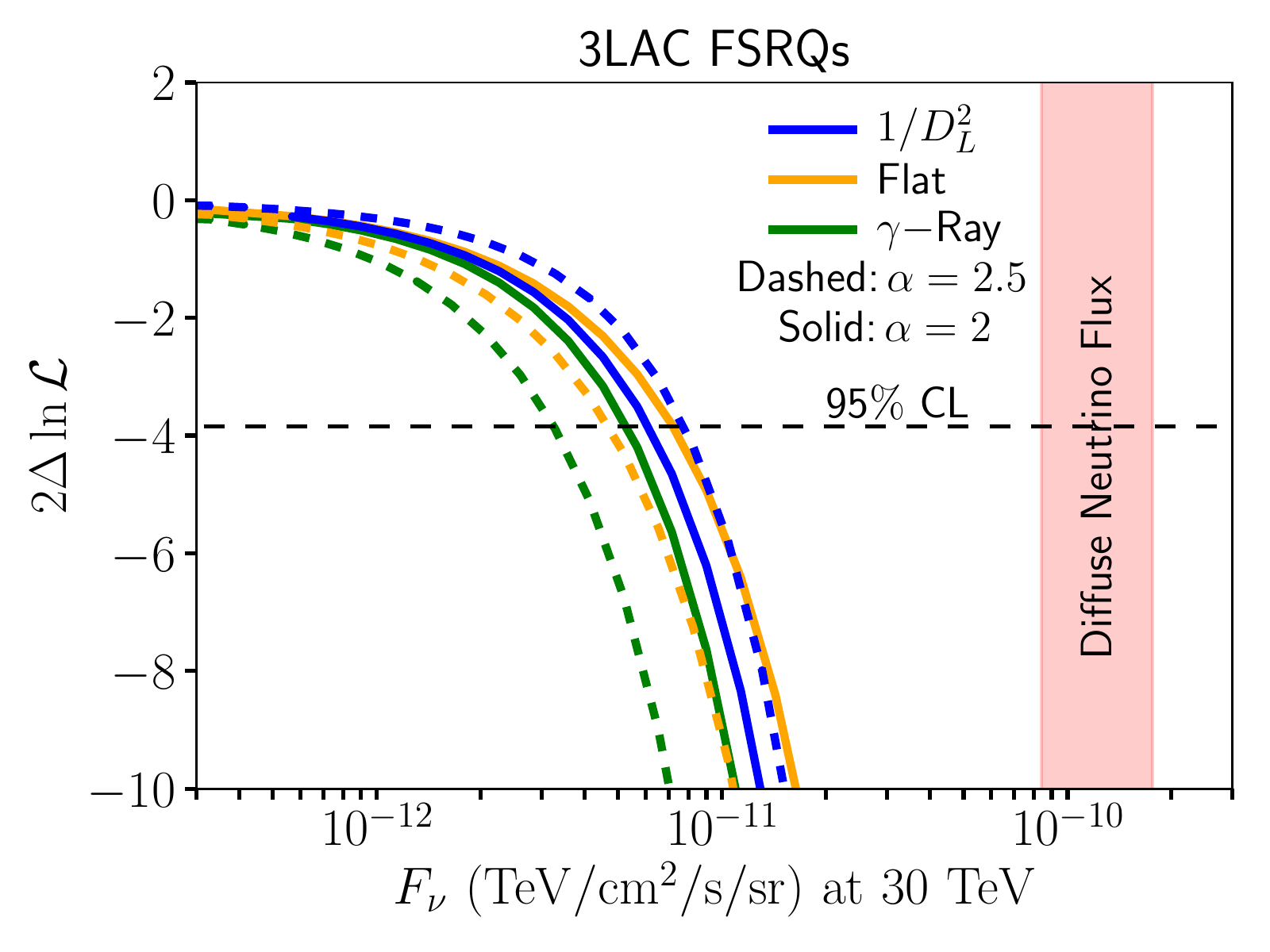} 
\includegraphics[width=3.0in,angle=0]{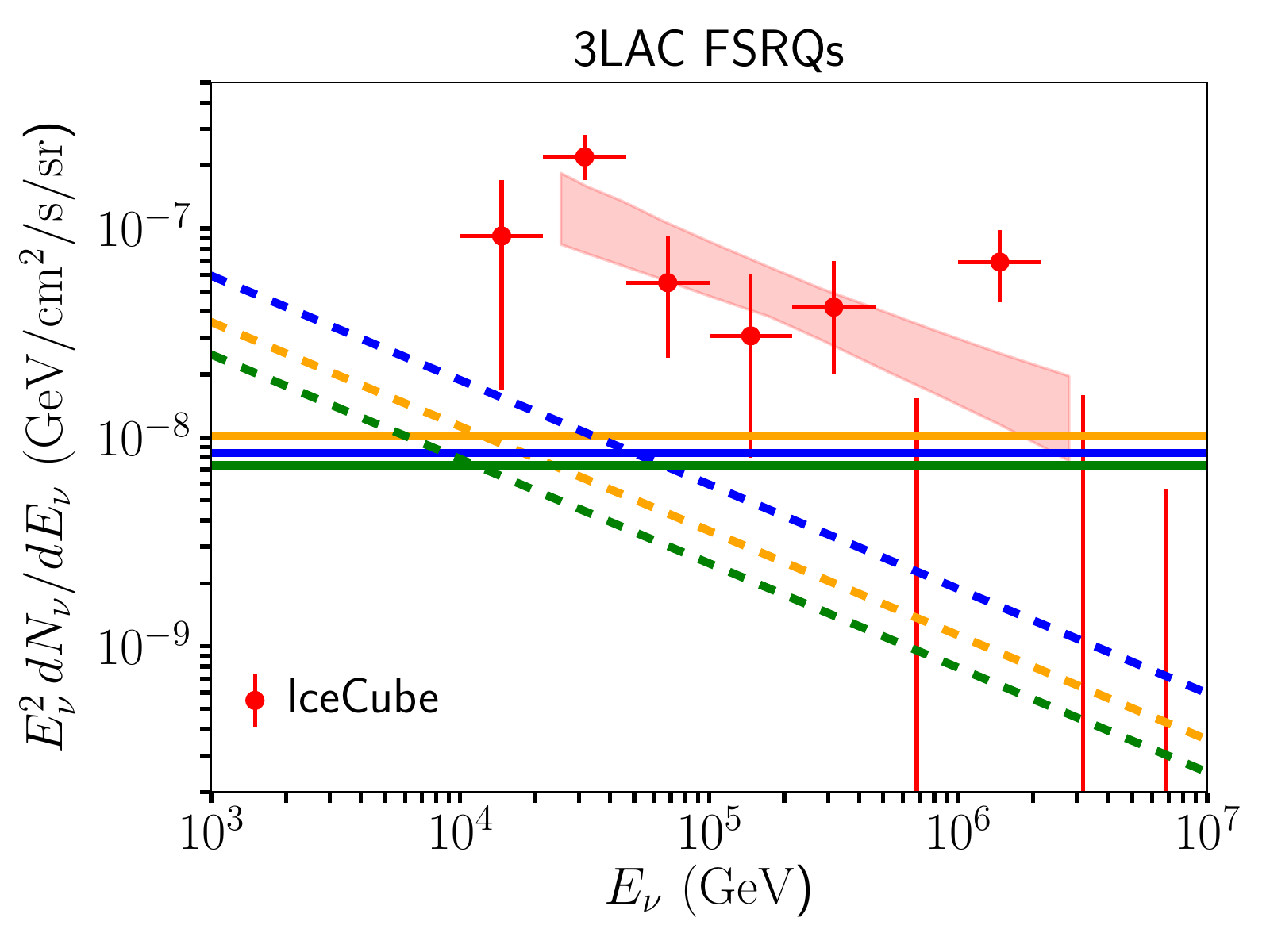}
\caption{As in Fig.~\ref{3LAC}, but for the case of blazars identified as BL Lac objects (top) or flat spectrum radio quasars (bottom). Our analysis identifies no evidence in favor of neutrino emission from these source populations and we are able to place strong limits on their contribution to IceCube's diffuse neutrino flux.}
\label{bllacfsrq}
\end{figure}

If there is no correlation between the gamma-ray and neutrino emission from AGN, increasing the value of $\delta$ will predict more bright blazars and thus, on average, result in stronger limits on the total neutrino flux from a given population. But if a relatively small fraction of the AGN in the sample are responsible for the bulk of the neutrino emission ($\delta > 0$), this likelihood function could acquire a larger value of $2\Delta \ln \mathcal{L}$ than those found with no such variations ($\delta=0$). In Fig.~\ref{TSsigma}, we show the results of our analysis for several values of $\delta$. We identify no evidence of neutrino emission from a subset of neutrino bright AGN within the membership of the 3LAC catalog.

Lastly, we repeat our analysis for the case of two subsets of the 3LAC catalog. In Fig.~\ref{bllacfsrq}, we show results for the blazars identified as BL Lacertae objects (top) and flat spectrum radio quasars (bottom). Once again, we find no evidence in favor of neutrino emission from these source populations and place strong limits on their contribution to IceCube's diffuse neutrino flux.

\section{Neutrinos From Blazar Flares}

\begin{figure}
\includegraphics[width=3.0in,angle=0]{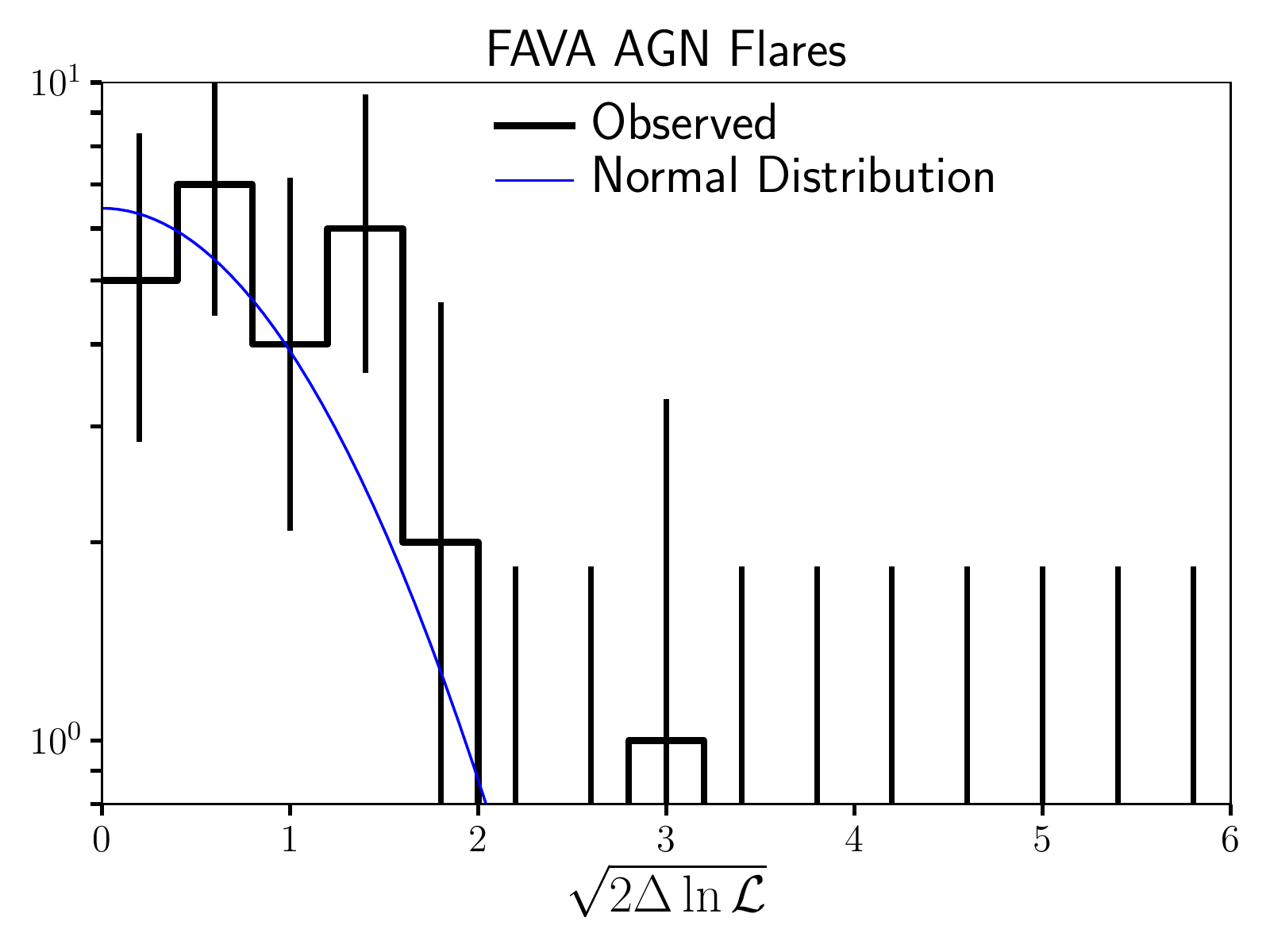}
\caption{The likelihood distribution in favor of neutrino point sources in the direction and time of the flaring sources cataloged in the Fermi All-Sky Varability Analysis (FAVA)~\cite{2017ApJ...846...34A}. The observed distribution shows no evidence of neutrino emission from these sources.}
\label{histflare}
\end{figure}

Motivated by the considerable variability in the gamma-ray and neutrino emission reported from TXS 0506+056, we consider in this section sources that produce neutrino emission during periods of gamma-ray flares~\cite{Liao:2018pta}. In particular, we consider the flaring periods of the sources described by the Fermi Collaboration's All-Sky Variability Analysis (FAVA)~\cite{2017ApJ...846...34A}. The FAVA catalog contains 95 sources that flared during the year covered by the IceCube dataset utilized in this study. For each of these sources, we calculated the likelihood in favor of neutrino emission in the direction of the source, using only events that arrived within the weeks reported as a flaring period. The likelihood distribution for these sources is shown in Fig.~\ref{histflare}. After accounting for trials, the most significant source ($2\Delta \ln \mathcal{L}$=10.2) provides evidence in favor of a neutrino flux at only the 1.5$\sigma$ level. We thus find no significant evidence that blazars produce neutrino emission while they are in a gamma-ray flaring state.

       

\section{Neutrinos Associated With Very-High Energy Fermi Events}

It is plausible that the extragalactic gamma-ray background as measured by Fermi could be dominated by leptonic processes at GeV energies, but becomes increasingly attributable to hadronic interactions at higher energies. If this is the case, then we should not expect the GeV emission from a given object to closely correlate with its neutrino emission. With this in mind, we perform a search for neutrino point sources in the directions of the $>$100 GeV Fermi events that were recorded during the period of time covered by the publicly available IceCube dataset. To avoid contamination from the Galactic Plane, we only utilize the 632 events with $|b|>10^{\circ}$.

\begin{figure}
\includegraphics[width=3.0in,angle=0]{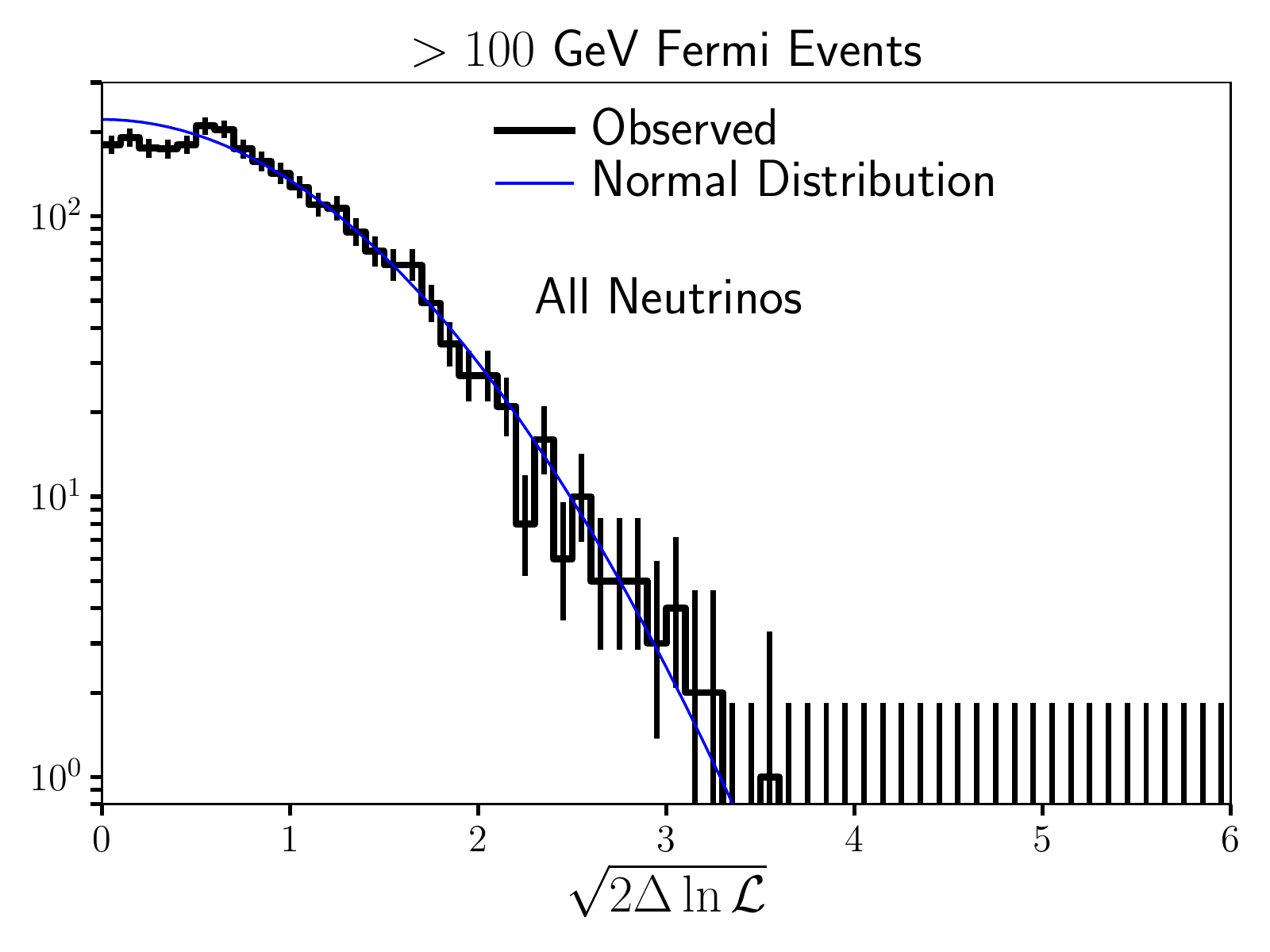}
\includegraphics[width=3.0in,angle=0]{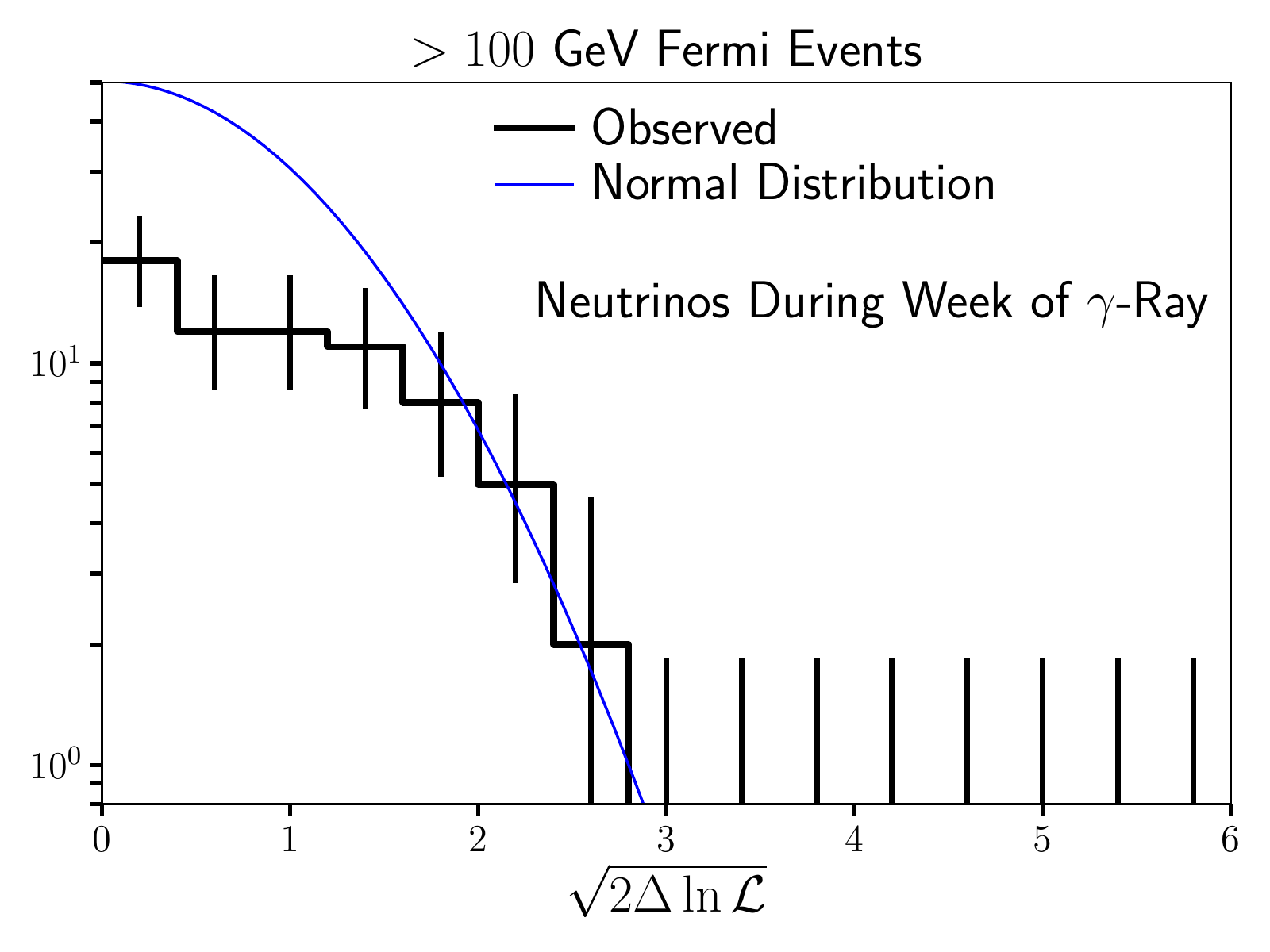}
\caption{The likelihood distribution in favor of a neutrino point source in the directions of Fermi events with $E_{\gamma} > 100$ GeV and with $|b|>10^{\circ}$. In the left frame, all such events are considered. In the right frame, we consider only those events that arrived within the same week of the very high-energy gamma ray. These distributions show no evidence of any neutrino point sources. We note that although the distribution shown in the right frame does not appear to be consistent with a normal distribution, this is due to the very small number of events that IceCube observes in a given direction over a time window of one week. A single neutrino from such a direction and time typically yields $\sqrt{2\Delta \ln \mathcal{L}} \sim 2-3$, explaining the dearth of positive but low values of the log-likelihood.}
\label{histFermi}
\end{figure}

In Fig.~\ref{histFermi}, we plot the likelihood distribution for these sources. In the left frame, we consider IceCube events from the entire dataset, while in the right frame we only utilize data from the week of the observed gamma ray. In neither case do we identify any evidence in favor of neutrino emission.

%

\section{Other Sources of High-Energy Neutrinos}        
      
The analyses carried out in the previous sections make it clear that the diffuse flux of high-energy neutrinos observed by the IceCube Collaboration cannot originate from a small number of very bright sources. In particular, very little of this neutrino flux (no more than $\simeq$\,5-15\%) could potentially originate from blazars. The majority of the astrophysical neutrinos observed by IceCube must instead be produced by a large number of comparatively faint sources. In this section we consider two sources classes that are each promising in this respect: non-blazar AGN and star-forming galaxies.

\subsection{Non-Blazar AGN}

In the unified model of AGN, blazars are those active galaxies with a jet aligned in the direction of the observer. Given that such jets are generally quite narrow, only a small fraction of AGN are observed as blazars, and there exists a much larger population of less luminous AGN with mis-aligned jets~\cite{Urry:1995mg}. 

It has been shown that Fermi's isotropic gamma-ray background (IGRB)~\cite{Ackermann:2014usa} is dominated at high-energies by emission from unresolved, non-blazar AGN~\cite{Hooper:2016gjy} (see also Refs.~\cite{Inoue:2011bm,DiMauro:2013xta}), along with smaller but non-negligible contributions from star-forming galaxies~\cite{Linden:2016fdd,Tamborra:2014xia,Ackermann:2012vca} and blazars~\cite{Cuoco:2012yf,Harding:2012gk,Ajello:2011zi,Ajello:2013lka,Stecker:2010di} (possibly among other sources, including galaxy clusters~\cite{Zandanel:2014pva}, millisecond pulsars~\cite{Calore:2014oga,Hooper:2013nhl}, and propagating ultra-high energy cosmic rays~\cite{Taylor:2015rla,Ahlers:2011sd}).\footnote{Although blazars generate more total high-energy gamma-ray emission than non-blazar AGN~\cite{TheFermi-LAT:2015ykq}, most of the emission from blazars has been resolved into individual sources. It is the unresolved, isotropic background that is dominated by the less luminous but more numerous non-blazar AGN population.} More quantitatively, Ref.~\cite{Hooper:2016gjy} concludes that non-blazar AGN account for $83.3^{+27.4}_{-10.1}\%$ of the photons above 1 GeV that make up the IGRB (see also Refs.~\cite{Fornasa:2015qua,DiMauro:2016cbj,DiMauro:2015tfa,Ajello:2015mfa,Cholis:2013ena,Cavadini:2011ig,Siegal-Gaskins:2013tga,Xia:2015wka,Cuoco:2015rfa,Shirasaki:2014noa,Shirasaki:2015nqp}).

It was shown in Ref.~\cite{Hooper:2016jls} that if the gamma-ray emission observed from non-blazar AGN is largely hadronic in nature, then one should expect these sources to also produce a diffuse flux of high-energy neutrinos that is qualitatively similar to that observed by IceCube. Despite being much fainter than individual blazars, the nearest and most luminous non-blazar AGN are predicted to produce potentially observable neutrino fluxes~\cite{Blanco:2017bgl}.

\begin{figure}
\includegraphics[width=3.0in,angle=0]{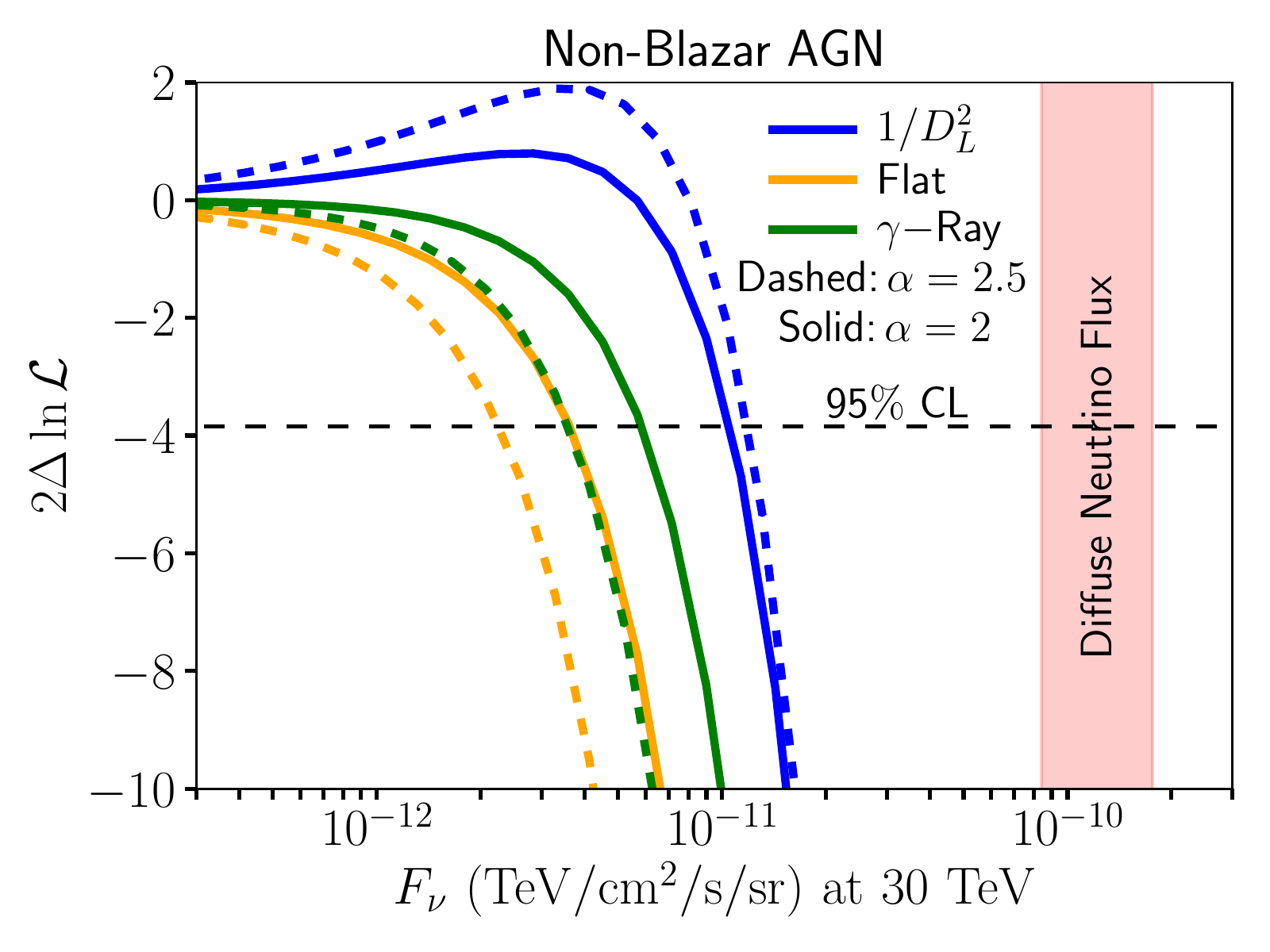}
\includegraphics[width=3.0in,angle=0]{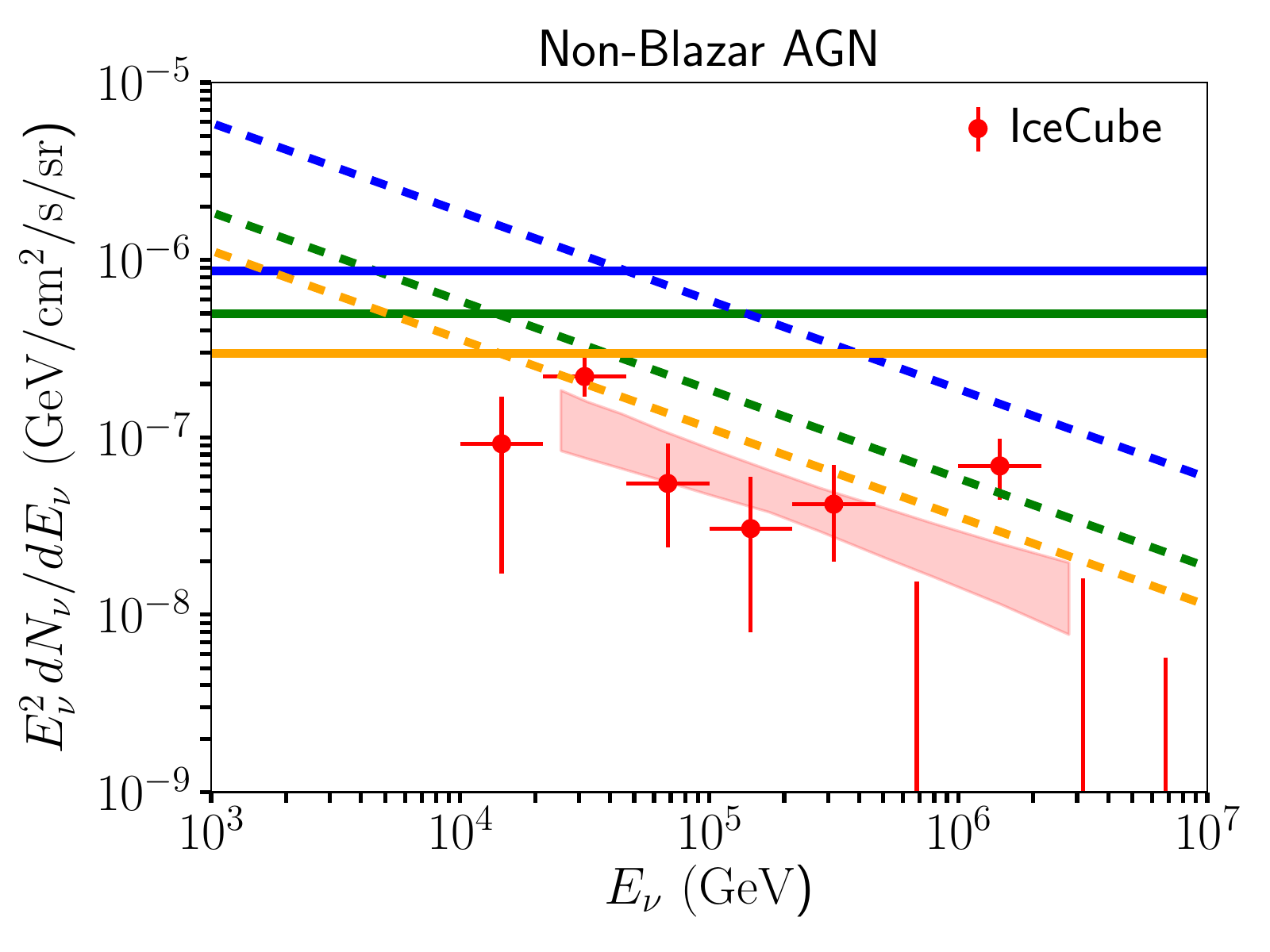}
\caption{As in Fig.~\ref{3LAC}, but for non-blazar AGN. Although our analysis identifies no significant evidence in favor of neutrino emission from these sources, the large number and low luminosities of this source population lead to a much weaker limit on their contribution to the total diffuse neutrino flux. Non-blazar AGN could plausibly produce the entirety of the diffuse neutrino flux reported by the IceCube Collaboration.}
\label{rdg}
\end{figure}

In Fig.~\ref{rdg}, we show the results of our search for neutrinos in the directions of non-blazar AGN. Due to their much lower luminosities, only a small number non-blazar AGN have been detected at gamma-ray wavelengths. For this reason, we utilize a collection of only 19 non-blazar AGN, including the 16 sources listed in Table~1 of Ref.~\cite{Hooper:2016gjy}, as well as 3FGL J0322.5-3721, 3FGL J0334.2+3915 (4C +39.12), and 3FGL J0758.7+3747 (NGC 2484). Due to their high degree of observed variability, we do not include the sources NGC 1275 or IC 310 in our population analysis, as their gamma-ray emission is likely to be dominated by leptonic processes.

The results shown in the left frame of Fig.~\ref{rdg} are qualitatively similar to those obtained in the case of blazars. In particular, we do not identify any significant evidence in favor of neutrinos from non-blazar AGN, and none of the 19 sources studied yielded any significant preference for neutrino emission (the largest evidence found was $2\Delta \ln \mathcal{L}=1.5$ for 3C 111). We also note that neither NGC 1275 nor IC 310 show any evidence of neutrino emisison.

In the right frame of this figure, we present the constraints on the total contribution to the diffuse neutrino flux from non-blazar AGN. This constraint is far weaker than those found for blazars due to the fact that the vast majority of non-blazar AGN have not yet been resolved as individual sources by Fermi. So instead of multiplying the limit on the flux by a modest factor of 1.4 to account for the unresolved population (as we did in the case of blazars), we account for the lack of completeness in the non-blazar AGN catalog as described in Ref.~\cite{Hooper:2016gjy}, corresponding to a completeness factor of $\simeq 90$. Once this is taken into account, we find that non-blazar AGN could very plausibly produce the entirety of the diffuse neutrino flux reported by the IceCube Collaboration.

\subsection{Star-Forming Galaxies}    
      
An even larger class of very low luminosity gamma-ray sources is that of star-forming galaxies~\cite{Ackermann:2012vca}. Despite being very faint individually, this source class produces much of the IGRB, in particular at GeV-scale energies~\cite{Linden:2016fdd,Fields:2010bw,Tamborra:2014xia}. If the gamma-ray emission observed from these sources is hadronic in origin, they too could contribute significantly to the diffuse flux of high-energy neutrinos~\cite{Loeb:2006tw}.

\begin{figure}
\includegraphics[width=3.0in,angle=0]{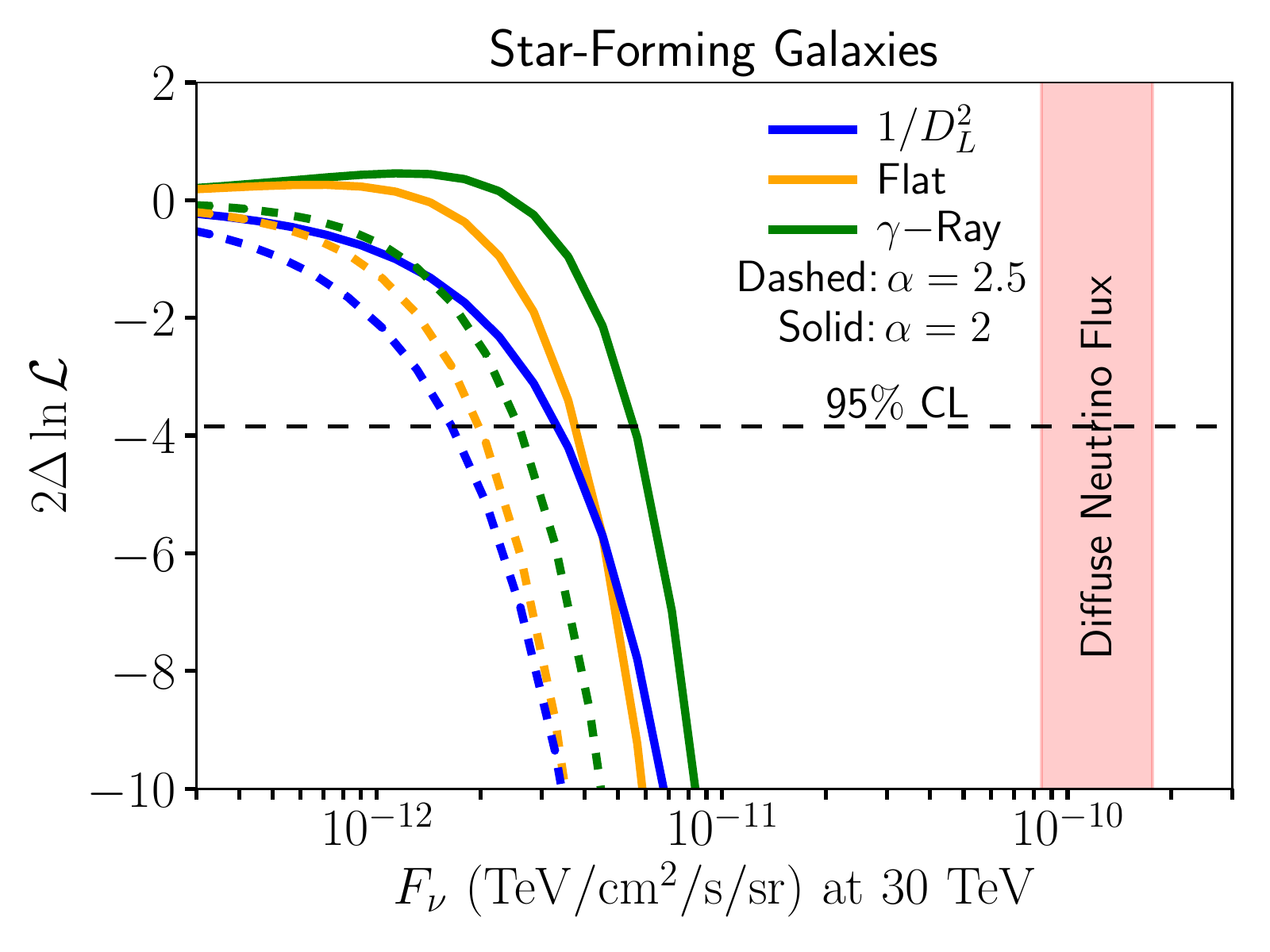}
\caption{As in Fig.~\ref{3LAC}, but for star-forming galaxies. As star-forming galaxies are much more numerous than AGN, our analysis cannot rule out the possibility that they could collectively produce a very large flux of high-energy neutrinos. That being said, spectral considerations indicate that star-forming galaxies cannot generate more than $\sim$10\% of the diffuse flux observed by IceCube without exceeding the measured intensity of the isotropic gamma-ray background~\cite{Bechtol:2015uqb}.}
\label{sfg}
\end{figure}

In Fig.~\ref{sfg}, we show the results of our search for neutrinos from the direction of five star-forming galaxies: NGC 253, NGC 1068, NGC 3034, NGC 4945 and Arp 220~\cite{Linden:2016fdd}. Once again, we find no statistically significant evidence of any neutrino emission (the greatest evidence was found for the case of NGC 1068, with $2\Delta \ln \mathcal{L}=2.7$). Not surprisingly, we find that these five star-forming galaxies do not produce a large fraction of the high-energy diffuse neutrino flux. However, because star-forming galaxies are so numerous, we cannot rule out the possibility that they might collectively produce very large fluxes of high-energy neutrinos. That being said, spectral considerations indicate that star-forming galaxies cannot generate the entirety of the diffuse flux observed by IceCube without exceeding the IGRB~\cite{Bechtol:2015uqb}, although this source class could still plausibly produce on the order of 10\% of the total observed flux.

\section{Summary and Discussion}
\label{summary}

It is a remarkable time for the field of neutrino astrophysics. The origin of IceCube's diffuse flux of high-energy neutrinos has energized the scientific community, and yet the origin of these particles remains a mystery. The observation of neutrinos from the blazar TXS 0506+056 would represent the first detection of an individual source of high-energy neutrinos. Such a discovery would carry considerable implications for the fields of neutrino astronomy, cosmic-ray physics and gamma-ray astronomy.

In this article, we have made use of the publicly available IceCube dataset in an effort to search for evidence of neutrino point sources. Our analysis included 1) a blind search across the sky, 2) searches in the directions of individual gamma-ray sources, 3) a search in the directions and times of individual gamma-ray flares, and 4) a search in the directions (or directions and times) of individual very high-energy gamma-rays. In none of these searches did we find compelling evidence in favor of any neutrino point sources, a result that is consistent with previous studies~\cite{Aartsen:2016oji,Aartsen:2016lir}.

When studying populations of astrophysical objects, we conclude that blazars cannot be responsible for more than 5-15\% of IceCube's total diffuse neutrino flux. Furthermore, we find no evidence that neutrinos are produced by blazars during periods with bright gamma-ray flares. From this perspective, we are forced to conclude that if TXS 0506+056 is, in fact, a source of high-energy neutrinos, it must be a fairly extreme outlier in terms of its total neutrino luminosity, and sources of this type cannot be responsible for a large fraction of the overall diffuse neutrino flux. This is consistent with the results of Ref.~\cite{Murase:2018iyl}, for example, which argues that blazar flares can be responsible for no more than $\simeq$1-10\% of IceCube's diffuse flux. We would like to emphasize, however, that the results presented here are not necessarily in conflict with IceCube's reported detections of TXS 0506+056. Although our analysis identified no evidence in favor of neutrinos from the direction of this source, the time period covered by the publicly available dataset (May of 2011 to May of 2012) does not cover either of the periods in which this source was detected by IceCube. Furthermore, when averaged over the 9.5 years of the total IceCube dataset, the neutrino flux reported from TXS 0506+056 represents only 1\% of the total diffuse flux measured by IceCube. The existence of an individual source with these characteristics thus remains consistent with the results presented here (as well as with the results of Refs.~\cite{Aartsen:2016oji,Aartsen:2016lir}).

The results of this paper have significant implications for the kinds of astrophysical objects that could be responsible for IceCube's diffuse flux. In particular, the lack of the detection of individual point sources (with the possible exception of TXS 0506+056) requires that the diffuse high-energy neutrino flux must be generated by a large population of very faint sources~\cite{Ahlers:2014ioa,Murase:2016gly}. From this perspective, non-blazar AGN and star-forming galaxies each stand out as particularly promising sources. In this study, we have searched for emission from known gamma-ray emitting non-blazar AGN and star-forming galaxies, finding no statistically significant evidence for such emission. But unlike blazars, we cannot rule out the possibility that either of these source classes produces the entirety of the diffuse flux reported by IceCube, due to the large number and low-luminosities of these individual sources. Furthermore, given that previous analyses have shown that star-forming galaxies cannot generate a large fraction of this neutrino flux without exceeding the isotropic gamma-ray background as measured by Fermi~\cite{Bechtol:2015uqb}, non-blazar AGN appear to be the most promising class of sources to produce the observed diffuse high-energy neutrino flux. Given that the limits we have placed on the diffuse neutrino flux from this source class (see the right frame of Fig.~\ref{rdg}) are generally within a factor of a few of the measured flux, the prospects are encouraging~\cite{Blanco:2017bgl} for the detection of these sources by IceCube or next generation neutrino telescopes~\cite{Aartsen:2014njl}.

\bigskip
\bigskip
\bigskip

\textbf{Acknowledgments.} We would like to thank Kohta Murase, Tyce DeYoung, Mauricio Bustamante, Naoko Neilson and Keith Bechtol for helpful discussions. DH is supported by the US Department of Energy under contract DE-FG02-13ER41958. Fermilab is operated by Fermi Research Alliance, LLC, under contract DE- AC02-07CH11359 with the US Department of Energy.

\bibliography{icecube}
\bibliographystyle{JHEP}

\end{document}